\documentclass[aps,groupedaddress]{revtex4}
\usepackage{graphicx}
\usepackage{subfig}
\usepackage{latexsym}

\begin{document}

\title{\bf{Entropy for theories with indefinite causal structure}}

\author{Sonia Markes}
\email{smarkes@perimeterinstitute.ca}
\affiliation{Department of Applied Mathematics, University of Waterloo, Waterloo, Ontario, Canada, N2L 3G1}
\affiliation{Perimeter Institute for Theoretical Physics, Waterloo, Ontario, Canada, N2L 2Y5}

\author{Lucien Hardy}
\email{lhardy@perimeterinstitute.ca}
\affiliation{Perimeter Institute for Theoretical Physics, Waterloo, Ontario, Canada, N2L 2Y5}

\begin{abstract}
Entropy is a concept that has traditionally been reliant on a definite notion of causality. However, without a definite notion of causality, the concept of entropy is not all lost. Indefinite causal structure results from combining probabilistic predictions and dynamical space-time. Combining the probabilistic nature of quantum theory and dynamical treatment space-time from general relativity is an approach to the problem of quantum gravity. The causaloid framework lays the mathematical groundwork to be able to treat indefinite causal structure. In this paper, we build on the causaloid mathematics and define a causally-unbiased entropy for an indefinite causal structure. In defining a causally-unbiased entropy, there comes about an emergent idea of causality in the form of a measure of causal connectedness, termed the Q factor.
\end{abstract}

\maketitle

\section{Introduction}
In Newtonian physics, physical processes are understood with respect to a fixed spatial coordinate system and a time parameter, which is absolute and ever increasing. Predictions are entirely deterministic. Quantum theory and general relativity depart from this classical picture in opposing manners. Quantum theory gives probabilistic predictions as to the outcomes of measurements, but retains fixed space and time coordinates. On the other hand, general relativity is deterministic, but shows that space and time form a dynamical structure. Reconciling these fundamental philosophical differences is one of the many challenges one is faced with in trying to construct a theory of quantum gravity. There have been many different approaches to this problem with many different results \cite{Penrose:1972ia, Hawking:1978jz, Rovelli:1989za, Thiemann:2007zz, Sorkin:1987cd, Ambjorn:2006jf}. One way of moving forward is to dismiss classical assumptions and create a probabilistic theory that has a dynamic causal structure. However, what results is indefinite casual structure. This is more radical than either probabilistic predictions or dynamical space-time structure. In general relativity, a separation between space-time locations is either space-like or time-like. An indefinite causal structure would allow for a separation between space-time locations to be something like a quantum superposition of a space-like and a time-like separation. While we may be uncertain of the causal structure of the path between measurements, we know where in space-time we make measurements, what measurements we have made, and what outcomes we get. With this data, we can examine probabilistic correlations for information. The causaloid framework (\cite{Hardy:2005fq},\cite{Hardy:2006uc},\cite{Hardy:2008fd}) provides us with the necessary structure. We will outline the essentials of this framework in Section 2.

It is natural in discussions of causal structure to raise the question of entropy. The second law of thermodynamics tells us that in an isolated system, entropy can increase or remain the same, but it can never decrease \cite{Landau&Lifshitz:1980}. In information theory, entropy is viewed as being a measure of uncertainty before we measure a state or equivalently, the amount of information gained by upon learning the state of a system \cite{Nielsen&Chuang:2000}. Inherent in both concepts of entropy is an assumed causal structure, specifically that there exists a background time. The standard definition of entropy is in the context of a definite causal structure with reference to absolute time. In order to make sense of entropy in an indefinite causal structure, a clear definition must be established. To do so requires consideration of the following questions:
\begin{quote}
\emph{What are the concepts from the usual picture of entropy in a definite causal structure that are necessary to define entropy? What are the analogues to these concepts in a picture with indefinite causal structure?}
\end{quote}
Using the formalism introduced in the causaloid framework, we are able to provide answers to these questions and then, define a causally-unbiased entropy.

In Section 2, we will review the relevant aspects of the causaloid framework. We then proceed with the new developments. In Section 3, we define a new type of product that is utilized in the work on entropy. The definition of causally-unbiased entropy and resulting features are developed in Section 4.

\section{Causaloid framework}
\subsection{The Picture}
Every experiment results in a set of data from making measurements on a system. Each piece of data could be thought of as a card with three pieces of information on it; where the measurement is made in space-time, what is measured, and what the result of the measurement is. We will represent each card (or piece of data) as $(x, f_x, y_x)$ where $x$ denotes the space-time information, $f_x$ denotes the information pertinent to a choice of measurement or action, and $y_x$ denotes the information regarding an observation or outcome of a measurement. The set of all possible cards (i.e. all possible measurements with all possible outcomes with every space-time configuration) is denoted $V$. We can imagine running an experiment an infinite number of times so as to be able to obtain relative frequencies. In order for the cards to tell us the relative frequencies, we must systematically sort them.

Each distinct $x$ is defined as an \emph{elementary region} of space-time. A \emph{composite region}, denoted $\mathcal{O}_1$, is a set of elementary regions. (Note: These definitions of ``elementary region" and ``composite region" differ from those in \cite{Hardy:2005fq,Hardy:2006uc,Hardy:2008fd}.) Therefore, these cards can be sorted according to their associated space-time region. The set of all possible cards with the same space-time information $x$ written on them is the \emph{measurement information for elementary region $x$}. We denote this set as $R_x$. The measurement information for composite region $\mathcal{O}_1$ is the union of all sets of measurement information for the elementary regions contained within the composite region. More concisely,
\begin{equation}
R_1 \equiv \bigcup_{x\in\mathcal{O}_1} R_x
\end{equation}
We can further sort the measurement information in a region. The \emph{procedure in a region}, denoted $F_x$, is the set of all distinct choices of measurement recorded for the region $x$.
\begin{equation}
F_x \equiv \bigcup_{all \hspace{1pt} f_x,y_x} \left(x, f_x, y_x\right)
\end{equation}
For composite regions, we define the following set:
\begin{equation}
F_1 \equiv \bigcup_{x\in\mathcal{O}_1} F_x
\end{equation}
Similarly, the \emph{outcome set in a region}, denoted $Y_x$ is defined to be the set of all distinct outcomes of a measurement recorded for the region $x$.
\begin{equation}
Y_x \equiv \bigcup_{all \hspace{1pt} y_x} \left(x, f_x, y_x\right)
\end{equation}
Again, for composite regions, we define
\begin{equation}
Y_1 \equiv \bigcup_{x\in\mathcal{O}_1} Y_x
\end{equation}
Notice that $Y_x \subseteq F_x \subseteq R_x$ and $Y_1 \subseteq F_1 \subseteq R_1$. The composite structure we expected of our space-time regions is reflected structure throughout these sets. A set of cards with the measurement information for a region has no more or less structure than an elementary region of space-time. Therefore, without adding structure or losing generality, we can take the sets $R_x$ to be elementary regions, at least, for the purposes of  this paper. From this point forward, the term \emph{region} will be used interchangeably to refer to objects of type $x$ or $\mathcal{O}_1$ and type $R_x$ or $R_1$.

Notice that the set of all cards $V$ can be viewed as all the cards from all (elementary) regions.
\begin{equation}
V = \bigcup_{all \hspace{1pt} x} R_x
\end{equation}
So $V$ is the largest of all regions that can be considered.

These definitions provide a firm foundation on which the causaloid framework rests both mathematically and conceptually.

\subsection{First level physical compression}
The most basic quantity that we would want to be able to calculate is the probability that a certain (set of) outcome(s) is observed given that a certain (set of) measurement(s) has been performed at a (set of) location(s) in space and time. Suppose that the set of locations we are interested in is $\mathcal{O}_1$. The set of all the cards corresponding to these locations called $R_1$. We write pairings of measurements and corresponding outcomes in $R_1$ as $(Y_1, F_1)$. A specific outcome and measurement pair is denoted as $\alpha_1$ (or equivalently, $(Y^{\alpha_1}_1, F^{\alpha_1}_1)$). The set of all $\alpha_1$ in region $\mathcal{O}_1$ is $\Upsilon_1$. The set comprised of all the cards not in $R_1$ is $V-R_1$. We call $(Y_{V-R_1}, F_{V-R_1})$ the \emph{generalized preparation} because it is the information that surrounds $R_1$ not only from the immediate past, but from the future and the rest of space-time as well. By the choices we make in setting up the experiment, we can put conditions on the generalized preparation such that ${\rm Prob}(Y_V | F_V)$ is well-defined. (See Ref.\cite{Hardy:2005fq} for details.) Then we can write
\begin{equation}
{\rm Prob}(Y_V | F_V) = {\rm Prob}( Y_1, Y_{V-R_1} | F_1, F_{V-R_1} )
\end{equation}
For a specific pair $\alpha_1 \Leftrightarrow (Y_1^{\alpha_1}, F_1^{\alpha_1})$, we can write this probability as
\begin{equation}
\label{p_alpha1 w gp}
{\rm Prob}( Y_1^{\alpha_1}, Y_{V-R_1} | F_1^{\alpha_1}, F_{V-R_1} )
\end{equation}
We use the short-hand $p_{\alpha_1}$ to denote the probability defined in Eq.(\ref{p_alpha1 w gp}). One way to specify the state of a system is to list all the possible $p_{\alpha_1}$ for elements of $R_1$.
\begin{equation}
\left( \begin{array}{c}
\vdots \\
p_{\alpha_1} \\
\vdots \end{array} \right)
\hspace{5mm}
\alpha_1\in\Upsilon_1
\end{equation}
However, this over-specifies the state. We do not usually need to know the probability of every outcome of every measurement in order to determine what the complete state of the system is. Physical theories tell us what relationships exist between variables and what constraints those relationships place on the variables of the system. These relationships and constraints can be used to determine a reduced set of probabilities from which all other probabilities can be represented. The reduced set of probabilities is defined such that any probability can be written as a linear combination of the probabilities in the reduced set. Let us denote the reduced or fiducial set in $R_1$ as $\Omega_1\subseteq\Upsilon_1$. This process of going from the set of all the probabilities to the smallest essential set we call \emph{first level physical compression}. This can be expressed as
\begin{equation}
\mathbf{p}= \left( \begin{array}{c}
\vdots \\
p_{l_1} \\
\vdots \end{array} \right)
\hspace{5mm}
l_1\in\Omega_1\subseteq\Upsilon_1 \end{equation}
such that
\begin{equation}
p_{\alpha_1} = \mathbf{r}_{\alpha_1} \cdot \mathbf{p}
\end{equation}
where $\mathbf{r}_{\alpha_1}$ encodes the physical compression and therefore, is determined by the details of the physical theory. We can define a \emph{decompression matrix}, $\Lambda_{\alpha_1}^{l_1}$ such that
\begin{equation}
\Lambda_{\alpha_1}^{l_1} \equiv \mathbf{r}_{\alpha_1} \big|_{l_1}
\end{equation}
where $\mathbf{r}_{\alpha_1} \big|_{l_1}$ means the $l_1$ component of $\mathbf{r}_{\alpha_1}$.

\begin{figure}[!h]
  \begin{center}
  \subfloat[]{\label{1region}\includegraphics[scale=0.3]{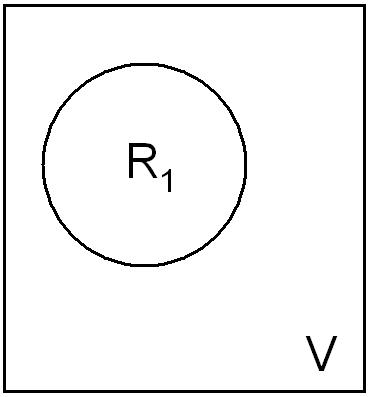}}
  \hspace{5pt}
  \subfloat[]{\label{2regions}\includegraphics[scale=0.3]{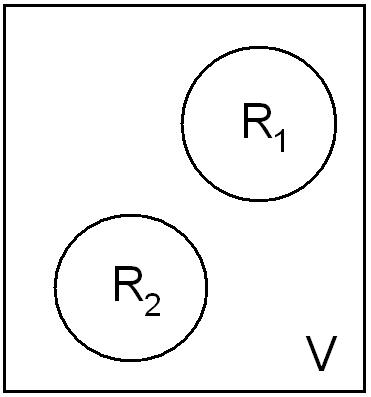}}
\end{center}
  \caption{(a) A single region $R_1$ in $V$  (b) Two regions $R_1$ and $R_2$ in $V$}
  \label{regionspic}
\end{figure}

\subsection{Second level physical compression and the causaloid product}
Let us consider two distinct regions $R_1, R_2 \subset V$. In a similar fashion to the single region case,
\begin{equation}
p_{\alpha_1\alpha_2}={\rm Prob}(Y_1^{\alpha_1},Y_2^{\alpha_2},Y_{V-R_1-R_2}|F_1^{\alpha_1},F_2^{\alpha_2},F_{V-R_1-R_2})
\end{equation}
We specify the state of the system by listing all $p_{\alpha_1\alpha_2}$.
\begin{equation}
\left( \begin{array}{c}
\vdots \\
p_{\alpha_1 \alpha_2} \\
\vdots \end{array} \right)
\hspace{5mm}
\alpha_1 \alpha_2 \in \Upsilon_1 \times \Upsilon_2
\end{equation}
where $\times$ is the cartesian product. It can be shown that
\begin{equation}
\label{1st level}
p_{\alpha_1\alpha_2}=\sum_{l_1l_2\in\Omega_1\times\Omega_2}\Lambda_{\alpha_1}^{l_1}\Lambda_{\alpha_2}^{l_2}p_{l_1l_2}
\end{equation}
which implies that the following list of probabilities is sufficient.
\begin{equation}
\left( \begin{array}{c}
\vdots \\
p_{l_1 l_2} \\
\vdots \end{array} \right)
\hspace{5mm}
l_1 l_2 \in \Omega_1\times\Omega_2\subseteq\Upsilon_1 \times \Upsilon_2
\end{equation}
This is effectively first level compression on each index. But if a physical theory has some connection between the two regions, $\Omega_1\times\Omega_2$ may no longer be the smallest set that is sufficient to represent all possible states. Then \emph{second level physical compression} is possible. It is defined to be
\begin{equation}
\mathbf{p} = \left( \begin{array}{c}
\vdots \\
p_{k_1 k_2} \\
\vdots \end{array} \right)
\hspace{5mm}
k_1 k_2 \in \Omega_{12}\subseteq\Omega_1\times\Omega_2\
\end{equation}
such that
\begin{equation}
\label{2nd level}
p_{\alpha_1\alpha_2}=\mathbf{r}_{\alpha_1\alpha_2}\cdot\mathbf{p}=\sum_{k_1k_2\in\Omega_{12}}\mathbf{r}_{\alpha_1\alpha_2}\big|_{k_1k_2}p_{k_1k_2}
\end{equation}
When $\Omega_{12}=\Omega_1\times\Omega_2$, second level compression is trivial. But it is proven in \cite{Hardy:2006uc} that it is possible that $\Omega_{12}\subset\Omega_1\times\Omega_2$.

Now we can define a second level decompression matrix. By comparing Eq.(\ref{1st level}) and Eq.(\ref{2nd level}), we infer that
\begin{equation}
\mathbf{r}_{\alpha_1\alpha_2}\big|_{k_1k_2} = \sum_{l_1l_2\in\Omega_1\times\Omega_2} \Lambda_{\alpha_1}^{l_1}\Lambda_{\alpha_2}^{l_2}\Lambda_{l_1l_2}^{k_1k_2}
\label{2nd level matrix}
\end{equation}
where
\begin{equation}
\Lambda_{l_1 l_2}^{k_1 k_2} = \mathbf{r}_{l_1l_2}\big|_{k_1k_2}
\end{equation}
which is the desired second level decompression matrix. This matrix encodes how we move from $p_{l_1l_2}$'s to $p_{k_1k_2}$'s. Using the definition of the first level decompression matrix, Eq.(\ref{2nd level matrix}) becomes
\begin{equation}
\mathbf{r}_{\alpha_1\alpha_2}\big|_{k_1k_2} = \sum_{l_1l_2\in\Omega_1\times\Omega_2} \Lambda_{l_1l_2}^{k_1k_2}\mathbf{r}_{\alpha_1}\big|_{l_1}\mathbf{r}_{\alpha_2}\big|_{l_2}
\end{equation}
This defines the \emph{causaloid product}, denoted $\mathbf{r}_{\alpha_1}\otimes^\Lambda\mathbf{r}_{\alpha_2}$ which unifies the different causal structure-specific products. Explicitly,
\begin{equation}
\mathbf{r}_{\alpha_1}\otimes^\Lambda\mathbf{r}_{\alpha_2}=\mathbf{r}_{\alpha_1\alpha_2}
\end{equation}
It is this product that allows us to look at the probabilistic correlations between arbitrary locations in space-time without specifying the causal relationship.

We have shown second level compression for the case where we have two regions. This is easily generalized for any number of regions. The object that would encode the compression for three regions would be $\Lambda_{l_1l_2l_3}^{k_1k_2k_3}$, for four regions would be $\Lambda_{l_1l_2l_3l_4}^{k_1k_2k_3k_4}$, etc. After second level compression over multiple regions, we have
\begin{equation}
\left( \begin{array}{c}
\vspace{5pt}
\Lambda_{\alpha_1}^{k_1}\\
\vspace{5pt}
\Lambda_{l_1l_2}^{k_1k_2}\\
\Lambda_{l_1l_2l_3}^{k_1k_2k_3}\\
\vdots \end{array} \right)
\end{equation}
There is a third level of physical compression that compresses these multi-region $\Lambda$-matrices to give the \emph{Causaloid}, $\mathbf{\Lambda}$, which is defined as
\begin{equation}
\mathbf{\Lambda}\equiv\left(\{\Lambda\}\mid\{\Lambda\}\subseteq\{\Lambda_{\alpha_1}^{k_1},\Lambda_{l_1l_2}^{k_1k_2},\ldots\}\right)
\end{equation}
where $\{\Lambda\}$ is determined by the rules of the physical theory (for detailed discussion of how this works see [2]).  By decompressing the set $\{\Lambda\}$, we can obtain the $\Lambda$-matrix for any set of regions. This means that the Causaloid gives us the ability to perform any calculation that the physical theory allows for.

\subsection{Well-defined probabilities}
Up to this point we have exclusively dealt with probabilities conditioned on procedures. It is more useful to also be able to condition on outcomes. Specifically, we'd like an expression for the following:
\begin{equation}
{\rm Prob}(Y_2^{\alpha_2} | Y_1^{\alpha_1}, F_1^{\alpha_1}, F_2^{\alpha_2})
\end{equation}
Using Bayes' Theorem, this becomes
\begin{equation}
{\rm Prob}(Y_2^{\alpha_2} | Y_1^{\alpha_1}, F_1^{\alpha_1}, F_2^{\alpha_2})
= \frac{{\rm Prob}(Y_1^{\alpha_1}, Y_2^{\alpha_2} | F_1^{\alpha_1}, F_2^{\alpha_2})}
{\sum_{Y_2^{\beta_2} \sim F_2^{\alpha_2}} {\rm Prob}(Y_1^{\alpha_1}, Y_2^{\alpha_2} | F_1^{\alpha_1}, F_2^{\alpha_2})}
\end{equation}
where $X_2^{\beta_2} \sim F_2^{\alpha_2}$ denotes that the sum is over all possible outcomes corresponding to the measurement $F_2^{\alpha_2}$ (in $R_2$). (For simplicity, we have suppressed the part of the notation denoting the generalized preparation.)
In the causaloid framework, this becomes
\begin{equation}
\label{probdef}
{\rm Prob}(Y_2^{\alpha_2}|Y_1^{\alpha_1},F_1^{\alpha_1},F_2^{\alpha_2})
= \frac{\mathbf{r}_{\alpha_1\alpha_2}\cdot\mathbf{p}} {\mathbf{r}_{\alpha_1\frown_2}\cdot\mathbf{p}}
\end{equation}
where $\mathbf{r}_{\alpha_1\frown_2}=\sum_{\beta_2}\mathbf{r}_{\alpha_1\beta_2}$. (The sum being over $\beta_2$ in this notation has the same meaning as the sum being over all outcomes consistent with $F_2$.) In order for this probability to be considered well-defined, the right hand side cannot depend on $V-R_1-R_2$. Since $\mathbf{r}_{\alpha_1\alpha_2}$ and $\mathbf{r}_{\alpha_1\frown_2}$ are determined exclusively by the physical theory, neither has any dependence on $V-R_1-R_2$. However, $\mathbf{p}$ does depend on $V-R_1-R_2$. This implies that in order for the probability Eq.(\ref{probdef}) to be well defined (i.e. not depend on $V-R_1-R_2$), it must vary with $\bf{p}$. The dependence on $\bf{p}$ can be removed altogether by requiring that $\mathbf{r}_{\alpha_1\alpha_2}$ be parallel to $\mathbf{r}_{\alpha_1\frown_2}$. Therefore, the above probability is well defined if and only if
\begin{equation}
\mathbf{r}_{\alpha_1\alpha_2}\parallel\mathbf{r}_{\alpha_1\frown_2}
\end{equation}
With this condition, we get
\begin{equation}
\label{parallel r's}
{\rm Prob}(Y_2^{\alpha_2}|Y_1^{\alpha_1},F_1^{\alpha_1},F_2^{\alpha_2}) = \frac{|\mathbf{r}_{\alpha_1\alpha_2}|}{|\mathbf{r}_{\alpha_1\frown_2}|}
\end{equation}

\section{$\odot^{\Gamma}$ product}
Consider two distinct regions; $R_A$ and $R_P$. By definition
\begin{eqnarray}
{\bf r}_{\alpha_A\alpha_P} = \mathbf{r}_{\alpha_A}\otimes^\Lambda\mathbf{r}_{\alpha_P} \nonumber \\
{\bf r}_{\beta_A\alpha_P} = \mathbf{r}_{\beta_A}\otimes^\Lambda\mathbf{r}_{\alpha_P} \nonumber
\end{eqnarray}
Suppose we wanted to take the dot product between two vectors of the above form. Using decompression matrices, we can write
\begin{eqnarray}
\label{devdot}
\lefteqn{{\bf r}_{\alpha_A\alpha_P}\cdot{\bf r}_{\beta_A\alpha_P}}\hspace{2cm}
&=& \left(\mathbf{r}_{\alpha_A}\otimes^\Lambda\mathbf{r}_{\alpha_P}\right)\cdot\left(\mathbf{r}_{\beta_A}\otimes^\Lambda\mathbf{r}_{\alpha_P}\right)\\
&=& \sum_{k_Ak_P} \left(\sum_{l_Al_P} \Lambda_{l_Al_P}^{k_Ak_P}\mathbf{r}_{\alpha_A}\big|_{l_A}\mathbf{r}_{\alpha_P}\big|_{l_P}\right) \left(\sum_{l'_Al'_P} \Lambda_{l'_Al'_P}^{k_Ak_P}\mathbf{r}_{\beta_A}\big|_{l'_A}\mathbf{r}_{\alpha_P}\big|_{l'_P}\right)\nonumber
\end{eqnarray}
where ${k_Ak_P}\in\Omega_{AP}$, ${l_Al_P}\in\Omega_A\times\Omega_P$, and ${l'_Al'_P}\in\Omega_A\times\Omega_P$. Notice that we can write $$\sum_{l_Al_P} \Lambda_{l_Al_P}^{k_Ak_P}\mathbf{r}_{\alpha_A}\big|_{l_A}\mathbf{r}_{\alpha_P}\big|_{l_P}$$ as $$\sum_{l_A\in\Omega_A}\left[\mathbf{r}_{\alpha_A}\big|_{l_A} \left(\sum_{l_P\in\Omega_P}\Lambda_{l_Al_P}^{k_Ak_P}\mathbf{r}_{\alpha_P}\big|_{l_P}\right) \right]$$
Similarly,
$$ \sum_{l'_Al'_P} \Lambda_{l'_Al'_P}^{k_Ak_P}\mathbf{r}_{\beta_A}\big|_{l'_A}\mathbf{r}_{\alpha_P}\big|_{l'_P}
= \sum_{l'_A\in\Omega_A}\left[\mathbf{r}_{\beta_A}\big|_{l'_A} \left(\sum_{l'_P\in\Omega_P}\Lambda_{l'_Al'_P}^{k_Ak_P}\mathbf{r}_{\alpha_P}\big|_{l'_P}\right) \right]$$
Define
$$\Gamma_{l_A}^{k_Ak_P}(\mathbf{r}_{\alpha_P}) \equiv \sum_{l_P\in\Omega_P}\Lambda_{l_Al_P}^{k_Ak_P}\mathbf{r}_{\alpha_P}\big|_{l_P}$$ and, similarly, $$\Gamma_{l'_A}^{k_Ak_P}(\mathbf{r}_{\alpha_P}) \equiv \sum_{l'_P\in\Omega_P}\Lambda_{l'_Al'_P}^{k_Ak_P}\mathbf{r}_{\alpha_P}\big|_{l'_P}$$
Using this, Eq.(\ref{devdot}) becomes
\begin{eqnarray}
\lefteqn{{\bf r}_{\alpha_A\alpha_P}\cdot{\bf r}_{\beta_A\alpha_P}}\hspace{2cm}
&=& \sum_{k_Ak_P} \left(\sum_{l_Al'_A} \Gamma_{l_A}^{k_Ak_P}({\bf r}_{\alpha_P})\Gamma_{l'_A}^{k_Ak_P}({\bf r}_{\alpha_P}) \mathbf{r}_{\alpha_A}\big|_{l_A}\mathbf{r}_{\beta_A}\big|_{l'_A} \right)
\end{eqnarray}
where $k_Ak_P\in\Omega_{AP}$ and $l_Al'_A\in\Omega_A\times\Omega_A$. This suggests that the essence of ${\bf r}_{\alpha_A\alpha_P}\cdot{\bf r}_{\beta_A\alpha_P}$ is a relationship between ${\bf r}_{\alpha_A}$ and ${\bf r}_{\beta_A}$ mediated by matrices that depend on ${\bf r}_{\alpha_P}$. Therefore, we can view Eq.(\ref{devdot}) as kind of product of ${\bf r}_{\alpha_A}$ and ${\bf r}_{\beta_A}$. Dot products of this form come up frequently enough that we will define this as the $\Gamma$-dot product and denote it as
\begin{equation}
{\bf r}_{\alpha_A} \odot^{\Gamma({\bf r}_{\alpha_P})} {\bf r}_{\beta_A} \equiv {\bf r}_{\alpha_A\alpha_P}\cdot{\bf r}_{\beta_A\alpha_P} = \left(\mathbf{r}_{\alpha_A}\otimes^\Lambda\mathbf{r}_{\alpha_P}\right)\cdot\left(\mathbf{r}_{\beta_A}\otimes^\Lambda\mathbf{r}_{\alpha_P}\right)
\label{gammadot}
\end{equation}
We will make use of this product later in the paper.

\section{Causally-unbiased entropy}
Standard definitions of entropy assume fixed causal structure. Here we develop a causally-unbiased definition of entropy in the causaloid formalism.

\begin{figure}[!h]
  \begin{center}
  \subfloat[]{\label{fixedcs}\includegraphics[scale=0.3]{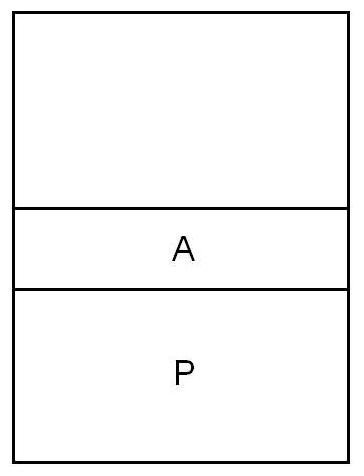}}
  \subfloat[]{\label{indefinitecs}\includegraphics[scale=0.3]{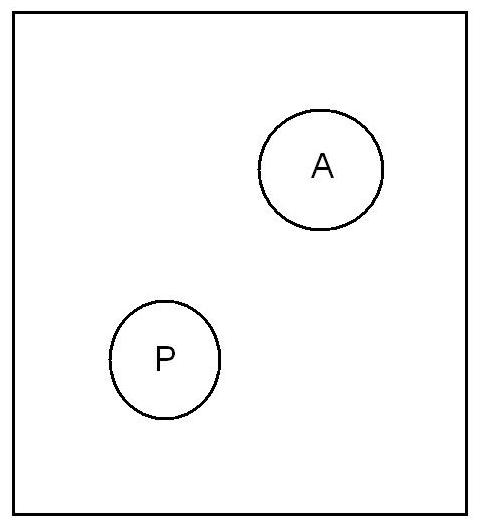}}
\end{center}
  \caption{(a) Fixed causal structure  (b) Indefinite causal structure}
  \label{cs}
\end{figure}

\subsection{The picture}
Shannon entropy \cite{Nielsen&Chuang:2000} for a classical state is defined as
\begin{equation}
S=-\sum_i p_i log_2 p_i
\end{equation}
The definition of $p_i$ used in this equation requires that the structure of space-time be organized with the following features:
\begin{itemize}
\item{a region of interest, $A$}
\item{an immediate past space-time region, $P$}
\item{sufficient data about what happened in $P$}
\item{a measurement $F_A$}
\item{a set of outcomes, $\{Y_A^i\}$, corresponding to $F_A$}
\end{itemize}
This allows us to write
\begin{equation}
p_i = {\rm Prob}(Y^i_A|F_A, {\rm data}_P)
\end{equation}
Removing all time bias from these features of space-time structure, we get
\begin{itemize}
\item{a region of interest, $A$}
\item{a reference region $P$}
\item{an outcome/measurement pair in $P$, $\{\alpha_P\}=\{(F_P,Y_P)\}$}
\item{a measurement $F_A$}
\item{a set of outcomes, $\{Y_A^i\}$, corresponding to $F_A$}
\end{itemize}
The reference region can be thought of as a kind of preparation region that is not limited to being in the causal past. In fact, the choice of reference region is arbitrary as illustrated in Fig. \ref{indefinitecs}.

The definition of $p_i$ in a causally-unbiased structure is
\begin{equation}
p_i = {\rm Prob}(Y_A^i|Y_P,F_P,F_A)
\end{equation}
(Since $P$ is arbitrary, we should technically say `$p_i$ with respect to the reference region $P$'. However, for the sake of brevity, we will assume that `with respect to $P$' is implied much as `with respect to the past' is taken as implied in the causally-biased situation.)

Using the above definition of $p_i$, we define the entropy relative to the reference data $(F_P,Y_P)$ as
\begin{equation}
\label{theo S}
S = - \sum_i {\rm Prob}(Y_A^i|Y_P,F_P,F_A) \log_2 \left({\rm Prob}(Y_A^i|Y_P,F_P,F_A)\right)
\end{equation}
Notice that this reduces to the causally-biased definition of entropy when $P$ is the past; $F_A$ measures the microstate in the classical case or measures in the basis where $\hat{\rho}$ is diagonal in the quantum case.

\subsection{In the causaloid framework}
Taking the probability to be well-defined, Eq.(\ref{parallel r's}) and Eq.(\ref{theo S}) give the following definition of entropy:
\begin{equation}
S = - \sum_{\alpha_A} \frac{|\mathbf{r}_{\alpha_A\alpha_P}|}{|\mathbf{r}_{\frown_A\alpha_P}|} \log_2 \left({\frac{|\mathbf{r}_{\alpha_A\alpha_P}|}{|\mathbf{r}_{\frown_A\alpha_P}|}}\right)
\end{equation}
Of course, this equation requires that $\mathbf{r}_{\alpha_A\alpha_P}\parallel\mathbf{r}_{\frown_A\alpha_P}$. Loosening this condition slightly, we can consider what happens when $\mathbf{r}_{\alpha_A\alpha_P}$ is nearly parallel to $\mathbf{r}_{\frown_A\alpha_P}$, using the definition of the probability from Eq.(\ref{probdef}). The entropy associated with this is
\begin{equation}
S^{\Lambda} = - \sum_{\alpha_A} \left( \frac{\mathbf{r}_{\alpha_A\alpha_P}\cdot\mathbf{p}}{\mathbf{r}_{\frown_A\alpha_P}\cdot\mathbf{p}} \right) \log_2 \left(\frac{\mathbf{r}_{\alpha_A\alpha_P}\cdot\mathbf{p}}{\mathbf{r}_{\frown_A\alpha_P}\cdot\mathbf{p}}\right)
\label{generalentropy}
\end{equation}
It becomes necessary to shorten the notation for the following work so $\mathbf{r}_{\alpha_A\alpha_P}$ will be denoted as $\mathbf{v}_i$ (where the index $\alpha_A$ is represented by $i$) and $\mathbf{r}_{\frown_A\alpha_P}$  will be denoted as $\mathbf{u}$. As with any vector, $\mathbf{v}_i$ can be decomposed into a component parallel to $\mathbf{u}$ and a component perpendicular to $\mathbf{u}$ (i.e. components in $\mathbf{\hat{u}}^{\parallel}$ and $\mathbf{\hat{u}}^{\perp}$, respectively). That is,
\begin{equation}
\mathbf{v}_i= v_i^{\parallel} \mathbf{\hat{u}}^{\parallel} +v_i^{\perp} \mathbf{\hat{u}}^{\perp}
\end{equation}
Using the unit vectors as defined, $\mathbf{p}$ can be decomposed as
\begin{equation}
\mathbf{p}= p_x \mathbf{\hat{u}}^{\parallel} + p_y \mathbf{\hat{u}}^{\perp} + \mathbf{p}^{\perp}
\end{equation}
where $\mathbf{p}^{\perp}$ is the component of $\mathbf{p}$ that is perpendicular to the plane defined by $\mathbf{u}$ and $\mathbf{v}_i$. The probability of interest, $p_i$, then becomes
\begin{eqnarray}
\lefteqn{p_i} \hspace{3pt}
&=& \frac{\mathbf{v}_i\cdot\mathbf{p}}{\mathbf{u}\cdot\mathbf{p}} \nonumber \\
&=& \frac{v_i^{\parallel}}{u}+k \frac{{v_i^{\perp}}}{u}
\label{notwdp}
\end{eqnarray}
where $k=\frac{p_y}{p_x}$. Notice that the first term is equivalent to a well-defined probability (Eq. \ref{parallel r's}). We require the second term to be small since the deviation from well-defined should be small. Since we have already required that $\mathbf{v}_i^{\perp}$ be small, we need only place restrictions on $k$.

\subsection{Bounds on $k$}
For the purposes of this subsection, we will work in the plane defined by $\mathbf{u}$ and $\mathbf{v}_i$. Define the angle between $\mathbf{u}$ and the projection of $\mathbf{p}$ into the plane to be $\theta$. Define the length of the projection of $\mathbf{p}$ into the plane to be $p_{xy}$. Using basic trigonometry, we get
\begin{eqnarray}
p_y &=& p_{xy} sin \theta \\
p_x &=& p_{xy} cos \theta
\end{eqnarray}
Therefore, k can be written in a form that is dependent on only one variable, as follows:
\begin{equation}
k = tan \theta
\end{equation}
As $\theta$ tends towards $\pm \frac{\pi}{2}$, $k$ tends to infinity. Therefore, to ensure that the second term of (\ref{notwdp}) is small, we require that $k$ be finite. Assume it to be a property of the state space for $\bf{p}$ that there exists some $0 < \theta_{max}$. Clearly, $|\theta_{max}| < \frac{\pi}{2}$ in order for $k$ to be finite. So $\theta$ is bounded as follows:
\begin{equation}
-\frac{\pi}{2} < -\theta_{max} \le \theta \le \theta_{max} < \frac{\pi}{2}
\end{equation}
The $k$ corresponding to $\theta_{max}$ will be denoted as $k_{max}$. Further bounds can be placed on $k$ by the state space of the physical theory. For our purposes, it is sufficient that $k$ is finite.

\subsection{$Q$ factor}
In light of (\ref{notwdp}), entropy, as defined in (\ref{generalentropy}), becomes
\begin{eqnarray}
\lefteqn{S^{\Lambda}}\hspace{7pt}
&=& -\sum_i \left(\frac{v_i^{\parallel}}{u}+k \frac{v_i^{\perp}}{u}\right) \log_2 \left(\frac{v_i^{\parallel}}{u}+k \frac{v_i^{\perp}}{u}\right) \nonumber\\
&=& -\sum_i \left(\frac{v_i^{\parallel}}{u}+k \frac{v_i^{\perp}}{u}\right) \left[\log_2\left(1+k\frac{v_i^{\perp}}{v_i^{\parallel}}\right)+\log_2\left(\frac{v_i^{\parallel}}{u}\right)\right]
\label{deventropy0}
\end{eqnarray}
Since $v_i^{\perp}$ is very small (as is implied by the fact that $\bf{v}_i$ and $\bf{u}$ are nearly parallel) and $k$ is finite, we can take a Taylor expansion (to leading order) of the first $\log_2$ term. Doing this gives
\begin{eqnarray}
\lefteqn{S^{\Lambda}}\hspace{7pt}
&=& -\sum_i \left(\frac{v_i^{\parallel}}{u}+k \left(\frac{{v_i^{\perp}}}{u}\right)\right) \left[ \frac{k}{\ln2} \left(\frac{{v_i^{\perp}}}{v_i^{\parallel}}\right) + {\cal O}\left({v_i^{\perp}}^2\right) + \log_2\left(\frac{v_i^{\parallel}}{u}\right) \right] \nonumber \\
&=& -\sum_i \left(\frac{v_i^{\parallel}}{u}\right) \log_2 \left(\frac{v_i^{\parallel}}{u}\right) + k\left(\frac{{v_i^{\perp}}}{u}\right) \log_2 \left(e\frac{v_i^{\parallel}}{u}\right) + {\cal O}\left({v_i^{\perp}}^2\right)
\end{eqnarray}
Notice that the first term is equivalent to the definition of entropy where $\bf{u} \parallel \bf{v}_i$ and that $S^{\Lambda}$ reduces to this definition when $v_i^{\perp}=0$. That is, when $\bf{u} \parallel \bf{v}_i$ (or equivalently, $v_i^{\perp}=0$)
\begin{equation}
S^{\Lambda} = S \equiv - \sum_{i} \left(\frac{v_i^{\parallel}}{u}\right) \log_2 \left(\frac{v_i^{\parallel}}{u}\right)
\end{equation}
For $v_i^{\perp} \neq 0$, we will define
\begin{equation}
Q = - \sum_i \left(\frac{{v_i^{\perp}}}{u}\right) \log_2 \left(e\frac{v_i^{\parallel}}{u}\right)
\end{equation}
Using $k_{max}$ as defined in the previous section, we can regard $k_{max} Q$ as a kind of correction to the causally-biased entropy. Then, to leading order
\begin{equation}
S - k_{max} Q \leq S^{\Lambda} \leq S + k_{max} Q
\end{equation}

\subsection{Understanding $Q$}
$Q$ is an entirely new quantity with no direct classical analogue so understanding its physical interpretation is a non-trivial matter. If we consider entropy as a measure of uncertainty, then $S$ is the measure of our uncertainty that the measurement $F_A$ in region $A$ will yield the specific outcome $Y_A^i$, given the data we have from the reference region $P$. Since our reference region $P$ is arbitrary, one way to view $Q$ is that it measures how completely the region $P$ ``prepares" region $A$. In this sense, preparation influences our uncertainty. In a definite causal structure, an immediate past region would completely prepare our region of interest and $Q$ would be zero. However, in the causally-indefinite picture, we cannot require a priori if the reference region that we have chosen will completely prepare our region of interest. If there are no influences on our uncertainty from outside region $P$, then the probability will be well-defined and $Q$ will be zero. But if there are influences on our uncertainty from outside region $P$, then the magnitude of $Q$ will reflect that.

\subsection{Using the $\odot^{\Gamma}$ product}
For the sake of completeness the $\bf u$'s and ${\bf v}_i$'s must be translated into ${\bf r}_{\frown_A\alpha_P}$'s and ${\bf r}_{\alpha_A\alpha_P}$'s. Notice that
\begin{eqnarray}
\frac{v_i^{\parallel}}{u} &=& \frac{{\bf v_i\cdot u}}{{\bf u\cdot u}} \\
\frac{{v_i^{\perp}}}{u} &=& \sqrt{\frac{v_i^2-{v_i^{\parallel}}^2}{u^2}} = \sqrt{\frac{\bf v_i\cdot v_i}{\bf u\cdot u} - \frac{({\bf v_i\cdot u})^2}{({\bf u\cdot u})^2}}
\end{eqnarray}
Substituting ${\bf r}_{\alpha_A\alpha_P}$ for ${\bf v}_i$ and ${\bf r}_{\frown_A\alpha_P}$ for $\bf u$ gives
\begin{eqnarray}
\frac{v_i^{\parallel}}{u}&=&\frac{{\bf r}_{\alpha_A\alpha_P}\cdot {\bf r}_{\frown_A\alpha_P}}{{\bf r}_{\frown_A\alpha_P}\cdot{\bf r}_{\frown_A\alpha_P}} \\
\frac{{v_i^{\perp}}}{u}&=& \sqrt{\frac{{\bf r}_{\alpha_A\alpha_P}\cdot {\bf r}_{\alpha_A\alpha_P}}{{\bf r}_{\frown_A\alpha_P}\cdot {\bf r}_{\frown_A\alpha_P}} - \frac{({\bf r}_{\alpha_A\alpha_P}\cdot {\bf r}_{\frown_A\alpha_P})^2}{({\bf r}_{\frown_A\alpha_P}\cdot {\bf r}_{\frown_A\alpha_P})^2}}
\end{eqnarray}
Using the $\Gamma$-dot product the above equations simplify to
\begin{eqnarray}
\frac{v_i^{\parallel}}{u}&=& \frac{{\bf r}_{\alpha_A} \odot^{\Gamma({\bf r}_{\alpha_P})} {\bf r}_{\frown_A}}{{\bf r}_{\frown_A} \odot^{\Gamma({\bf r}_{\alpha_P})} {\bf r}_{\frown_A}} \\
\frac{{v_i^{\perp}}}{u}&=& \sqrt{\frac{{\bf r}_{\alpha_A} \odot^{\Gamma({\bf r}_{\alpha_P})} {\bf r}_{\alpha_A}}{{\bf r}_{\frown_A} \odot^{\Gamma({\bf r}_{\alpha_P})} {\bf r}_{\frown_A}} - \left(\frac{{\bf r}_{\alpha_A} \odot^{\Gamma({\bf r}_{\alpha_P})} {\bf r}_{\frown_A}}{{\bf r}_{\frown_A} \odot^{\Gamma({\bf r}_{\alpha_P})} {\bf r}_{\frown_A}}\right)^2}
\end{eqnarray}
This allows us to completely specify the entropy of $R_A$ relative to a preparation $R_P$ in the causaloid framework. It is straightforward to generalize this to define the joint entropy of $R_A$ and $R_B$ with reference to a ``preparation" $R_P$. Simply redefine $\bf{u}$ and $\bf{v}_{ij}$ as
\begin{eqnarray}
{\bf v}_{ij} &=& {\bf r}_{\alpha_A\alpha_B\alpha_P} \\
{\bf u} &=& {\bf r}_{\frown_A\frown_B\alpha_P}
\end{eqnarray}
where $${\bf r}_{\alpha_A\alpha_B\alpha_P} = {\bf r}_{\alpha_A} \otimes^\Lambda {\bf r}_{\alpha_B} \otimes^\Lambda {\bf r}_{\alpha_P}$$ and $${\bf r}_{\frown_A\frown_B\alpha_P} = \sum_{\beta_A}{\bf r}_{\beta_A} \otimes^\Lambda \sum_{\beta_B}{\bf r}_{\beta_B} \otimes^\Lambda {\bf r}_{\alpha_P}$$
Using the same procedure as for one region, we get
\begin{eqnarray}
\frac{v_{ij}^{\parallel}}{u}&=& \frac{{\bf r}_{\alpha_A\alpha_B} \odot^{\Gamma({\bf r}_{\alpha_P})} {\bf r}_{\frown_A\frown_B}}{{\bf r}_{\frown_A\frown_B} \odot^{\Gamma({\bf r}_{\alpha_P})} {\bf r}_{\frown_A\frown_B}} \\
\frac{{v_{ij}^{\perp}}}{u}&=& \sqrt{\frac{{\bf r}_{\alpha_A\alpha_B} \odot^{\Gamma({\bf r}_{\alpha_P})} {\bf r}_{\alpha_A\alpha_B}}{{\bf r}_{\frown_A\frown_B} \odot^{\Gamma({\bf r}_{\alpha_P})} {\bf r}_{\frown_A\frown_B}} - \left(\frac{{\bf r}_{\alpha_A\alpha_B} \odot^{\Gamma({\bf r}_{\alpha_P})} {\bf r}_{\frown_A\frown_B}}{{\bf r}_{\frown_A\frown_B} \odot^{\Gamma({\bf r}_{\alpha_P})} {\bf r}_{\frown_A\frown_B}}\right)^2}
\end{eqnarray}
In this manner, we can define causally-unbiased entropy in the causaloid framework for any number of regions.

\section{Conclusions}
In a definite causal structure, the only thing required for a definition of entropy that is not in an indefinite causal structure is an immediate past region. Since there is no reason in an indefinite causal structure to choose any reference region over any other, we simply choose an arbitrary region. This ensures that we do not hold on to any pre-conceived notions of space-time and its connection to causality. The definition of the causally-unbiased entropy resulted in a correction to the causally-biased definition of entropy. In a sense, the Q factor gives us an emergent idea of causality. It is a measure of the extent to which our region of interest is causally connected to our reference (or ``preparation") region. If it is zero, the traditional ideas of causality are recovered. The next step would be determining how the Q factor could potentially be physically observed. To do so may require us to know more of the theoretical and mathematical properties of Q. Which mathematical properties of Shannon entropy hold for causally-unbiased entropy? What is the status of the Second Law of Thermodynamics in an indefinite causal structure? To go about answering this, we could consider how $S^{\Lambda}$ ``evolves'' along tubes through indefinite space-times. These questions will be the subjects of continuing work in the near future.

\section{Acknowledgements}
This work was supported by OGS. Research at Perimeter Institute for Theoretical Physics is supported in part by the Government of Canada through NSERC and by the Province of Ontario through MRI.


\begin{thebibliography}{99}

%\cite{Penrose:1972ia}
\bibitem{Penrose:1972ia}
  R.~Penrose and M.~A.~H.~MacCallum,
  ``Twistor theory: An Approach to the quantization of fields and space-time,''
  Phys.\ Rept.\  {\bf 6}, 241 (1972).
  %%CITATION = PRPLC,6,241;%%

%\cite{Hawking:1978jz}
\bibitem{Hawking:1978jz}
  S.~W.~Hawking,
  ``Quantum Gravity And Path Integrals,''
  Phys.\ Rev.\  D {\bf 18}, 1747 (1978).
  %%CITATION = PHRVA,D18,1747;%%

%\cite{Rovelli:1989za}
\bibitem{Rovelli:1989za}
  C.~Rovelli and L.~Smolin,
  ``Loop Space Representation of Quantum General Relativity,''
  Nucl.\ Phys.\  B {\bf 331}, 80 (1990).
  %%CITATION = NUPHA,B331,80;%%


%\cite{Thiemann:2007zz}
\bibitem{Thiemann:2007zz}
  T.~Thiemann,
  ``Modern canonical quantum general relativity,''
%\href{http://www.slac.stanford.edu/spires/find/hep/www?irn=7656084}{SPIRES entry}
  Cambridge, UK: Cambridge Univ. Pr. 819 (2007).


%\cite{Sorkin:1987cd}
\bibitem{Sorkin:1987cd}
  R.~D.~Sorkin,
  ``On the role of time in the sum over histories framework for gravity,''
  Int.\ J.\ Theor.\ Phys.\  {\bf 33}, 523 (1994).
  %%CITATION = IJTPB,33,523;%%

%\cite{Ambjorn:2006jf}
\bibitem{Ambjorn:2006jf}
  J.~Ambjorn, J.~Jurkiewicz and R.~Loll,
  ``Quantum gravity, or the art of building spacetime,''
  hep-th/0604212.
  %%CITATION = HEP-TH/0604212;%%

%\cite{Hardy:2005fq}
\bibitem{Hardy:2005fq}
  L.~Hardy,
  ``Probability theories with dynamic causal structure: A new framework for
  quantum gravity,''
  gr-qc/0509120.
  %%CITATION = GR-QC/0509120;%%

%\cite{Hardy:2006uc}
\bibitem{Hardy:2006uc}
  L.~Hardy,
  ``Towards quantum gravity: A framework for probabilistic theories with
  non-fixed causal structure,''
  J.\ Phys.\ A  {\bf 40}, 3081 (2007),
  gr-qc/0608043.
  %%CITATION = JPAGB,A40,3081;%%

%\cite{Hardy:2008fd}
\bibitem{Hardy:2008fd}
  L.~Hardy,
  ``Formalism Locality in Quantum Theory and Quantum Gravity,''
  gr-qc/0804.0054.
  %%CITATION = ARXIV:0804.0054;%%

\bibitem{Landau&Lifshitz:1980}
    L. D. Landau and E. M. Lifshitz, \textit{Course of Theoretical Physics Vol. 5: Statistal Physics Pt. 1} 3rd Ed. (Butterworth Heinemann, Oxford, 1980).

\bibitem{Nielsen&Chuang:2000}
  M. A. Nielsen and I. L. Chuang, \textit{Quantum Computation and Quantum Information} (Cambridge University Press, Cambridge, 2000).

\end{thebibliography}
\end{document}